# Alternate methods of evaluation for web sites concordant to IAS/IFRS Standards

Bostan I., Mates D., Grosu V., Iancu E.

**Abstract**— This work has as the principal theme, the study, analysis and implementation of the methodology for use the web sites in e-commerce. The authors try to deal with particular methodological and applied aspects inherent in the analysis of data from the interaction of man-Internet (Web-mining). The research methodology of this work will be focused on a prevalent optic multidisciplinary research based on the pillars of data mining and Web mining. The explosion of Internet and electronic commerce has made the most of business to have its own website. A company may engage internal costs for the development and operation of their website. The website can be designed for internal access (in which case it can be used for presentation and data storage company policies with references of customers) or for external access (they are created and used for promotional and advertising products and services company). The objective of this research, primarily concerns the definition of a repertoire of tools in analyzing e-business through the development process for web-usage mining; 2nd objective is oriented to management, recognizing and evaluating the web-sites in accountancy, as property intangible, which is a special case and very little studied in economic literature financial specialty, the authors try to achieve a national and international accounting treatment of the creation and development of web-sites. This paper is intended to have a predominantly interdisciplinary character, trying to connect decision theory viewpoint of knowledge, the theme of the interaction between humans and Internet and particularly in evaluating their in accounting firm. We chose this theme in an effort to tell us more about electronic commerce, especially the way this activity can be managed, evaluated and accounted, considering its importance for both the dealer (who is trying to promote and sell their products using a minimum cost), but also for the client (which, besides the fact that they can save time, have the opportunity and price comparison quality products on the market).

—————————— ◆ ——————————

## 1 INTRODUCTION

The electronic commerce is one of EU priorities, drafted for this purpose the electronic commerce Directive (EC Directive 2000/31/EC of June 8, 2000). This directive aims to redress certainty in e-commerce law, to improve confidence subscribers and Internet users. In this connection was established a stable legal framework on information society services, the requirements of internal market (free movement and freedom of establishment), whilst providing some measure of harmonization.

This directive aims primarily to establish a coherent European legal framework for e-commerce-watching support provided to avoid an excessive number of rules, based on internal market freedoms, taking into account the commercial realities and guaranteeing an effective control of objectives interest, based on the will to eliminate existing disparities in Justice of the Member States in how to establish a certainty able to foster greater consumer confidence and the companies.

This directive applies exclusively to service providers and information companies (e.g. the site internet operators), respecting the national provisions in force in Member States or which are set (usually the country of origin or "internal market clause), defining the place of establishment the provider to place an operating effectively and on undetermined time exercise an economic activity through stable arrangements. This rule of the country of origin, are fundamental elements of the Directive in order to create the clarity and certainty in the law necessary to enable service providers to propose their services across the EU.

The commercial communication must be clearly identifiable (Article 6), in order to recover the consumer confidence and ensure a real commercial practice. In addition, commercial communication via electronic messages must be clearly identified at the time the recipient receives, Member States should adopt provisions necessary for providers to send by email, unsolicited communications to consult records adverse commercial (eight out ), where they can register as individuals do not wish to receive such commercial communications.

In general, the directive adoption in member countries was satisfactory excepting Nordic countries who have decided to resort to adopting a law horizontal e-

- *Bostan I. is with the University of Suceava, 13 University Str., Suceava, Romania.*
- *Mateş D. is with the* ***West*** *University of Timişoara, 16 Pezzdalozi Str., Timişoara, Romania.*
- *GrosuV. is with the University of Suceava, 13 University Str., Suceava, Romania.*
- *Iancu E. is is with University of Suceava, 13 University Str., Suceava, Romania.*

commerce module to provide a national framework as clear as possible.

Electronic commerce is a topic of interest in the business world around the world as a way to promote products and business development. Today, we live in a society where information technology has advanced, the computer has become an indispensable necessity of communication, information and promote business, we can say that the business world, unless you website does not exist.

As the technology is more easily adopted, will be more and more retailers and customers will share the idea of e-commerce.

Some authors consider the Internet as an ideal business environment in which the small and medium businesses have the opportunity to successfully compete with large multinational companies. The research undertaken shows that these benefits are not so obvious for small and medium-sized and that ultimately, they can be operated only by the choice and implementation of optimal marketing strategies.

In a study published, Ernst & Young shows that online sales in 1999 reached the figure of 12-15 billion. Online buyers were at a rate of 54% female, 58% married, 58% aged between 30 and 49 years, 23% over 50 years, 19% between 18 and 29 years, 61% have children and 58% have incomes of between 30,000 and $ 69,000 per year. These figures refer to users in North America because America is where the Internet emerged and still remains the leader in the field. The analysis of data provided by Ernst & Young, can come off that segment of the population which is addressed to the Internet and online sales, is the high proportion of women, paradoxically, as is well known concern for the act of buying sex itself and evaluate the product according to its qualities perceptible by the sense organs work is not of the Internet. Large percentage (58%), recorded in the use of online commerce and the average age, explained by the time saved by resorting to this form to purchase. Another paradox is the greater percentage (23% vs. 19%) of those over 50 years using the Internet for commercial purposes, to the young. One explanation may be convenience of buying in this way, without having to move to the place of purchase. The global electronic commerce is in adulthood, internal stability, passing several major crises (the collapse of the dot-com sites unjustified and multi crises arising from loss / theft of millions of full data card), try to change their appearance through new standards (possibly future) Universal (3D Secure, Verified by Visa, MasterCard Secure Code, Payment Card Industry Data Security Standard, Contact Less Technology, Smart Cards, etc.) and tends towards efficiency and normality.

Of scientific viewpoint, the electronic commerce is defined as "a modern technology to do business, that addresses the needs of organizations, businesses and consumers to reduce transaction costs along with improving the quality of goods and services and increase the speed of delivery. The term can be used when using computer networks to search and retrieve informa-tion's to support human and institutional decision [1].

On the pragmatic viewpoint manifests importance and other concepts and ways to define. Among these, special interest shows the following:

♦ "all activities performed online, in order to arouse the interest of consumers before the sale and provide after sales support to consumers, given by Internet Computing magazine;

♦ "commercial transactions taking place in open networks, given by the European Organization for Trade and Development;

♦ "doing business online. This includes buying products via on-line services and Internet and electronic data exchange, where an institution's computer informing the computer purchase and transmit orders to other companies "given the Computer Desktop Encyclopedia;

♦ "electronic commerce technologies include all forms of electronic trading, sending electronic messages, electronic data exchange (EDI), electronic funds transfer (EFT), electronic mail, electronic catalogs, data bases on electronic electronic news and information services, payroll, electronic and other forms of electronic communication (FCE), access online services through the Internet and other forms of electronic data transmission for commercial purposes "by" Electronic Commerce for Small to Medium Sized Enterprises the University Monash in Australia.

♦ "electronic commerce is achieving electronic commercial activities. It is based on processing and transmission of data including text, sound and images. It includes various activities, including electronic trading products and services, network delivery of goods, electronic transfer of capital, trade in shares, the electronic transmission of dispatch sheets, direct marketing and service warranty and post - guarantee. Involves both products (household appliances, medical) and services (information, financial, legal) traditional activities (health, education) and new activities (virtual malls) "given by the European Commission, 1997[3];

- the commerce term its used to describe the information infrastructure to achieve the following functions:

- bringing to market products (example: cybermarketingul);

- meeting buyers with sellers (example: electronic trade, virtual stores, electronic cash transfer);

- payment of obligations to the state (electronic collection of fees and taxes);

- delivery of electronic goods [2].

***Occurrence and trends***

The electronic commerce has been defined by IBM in the first half of the 1990s, the most important component of the term e-business, and thereafter the electronic business scope to include many other areas later. Initially e-commerce (name that has since evolved into the form: ecommerce, like all new names in the digital age, which is formed by adding the letter s to the word old, but no longer use the hyphen) means advertising and trade made through the Internet. And as the inter-net age time is running very fast and the terms and technologies quickly become 'old', currently the term used to replace e-ecommerce is e-business. Electronic commerce [4] is an integrative concept, designating a variety of services: electronic catalogs, support systems for trade in goods and services, acquisition support systems for command, logistics and trading, statistical reporting systems and information management. Also, electronic commerce and refers to specific business activities, especially using electronic means (computer network) in a fully automated exchange of business.

I believe however that the latter sphere is actually scope's of e-business, e-commerce and e-business yet words are often considered synonymous, in many works that deal with this issue [8].

Information circulating in e-commerce directly between agencies involved (seller, buyer, bank, carrier, service agent), without using the paper support, printer or fax.

The electronic commerce entails the productivity across the economy, encourage trade both in goods and services and investment, create new sectors of activity, new forms of marketing and selling, new revenue streams and, not least, new jobs. Moreover, the existing model of e-commerce financial transaction, the Internet has become a channel of commerce with an undeniable power to facilitate and increase sales in a range of increasingly large products and services.

The term e-commerce includes also the business conducted on the network website. And most important asset of electronic commerce is the short time for the completion of a business; feature only offered by the Internet.

Examining the applications we have identified the following business models in electronic commerce: electronic stores, electronic supplies, electronic stores, virtual communities, providers of brokerage services.

The e-shop is managed by a company for marketing and selling their own products or services. Minimal catalog of products / services, market and technical descriptions for each position of the catalog. Average variable includes facilities for taking orders (by email or online forms), and expanded version includes the possibility of online payment (credit card or other electronic means). The main motivation of creating e-shops is attracting more customers without distance longer an impediment. This is the shortest path to a global presence to a company. The gains come from cost reduction and sales promotion, as well as increase sales[1].

*Electronic Supply* - For the procurement of goods and services, the large companies and public authorities hold auctions. The web publishing tender specifications decreases both time and cost of transmission is still more important increase in the number of companies that take timely informed about the auction, leading ultimately to increase competition and therefore lower prices.

*Electronic store* - universal store, is a collection of electronic shops, gathered under a common umbrella, for example a well known brand. Generally accepted a common payment method, guaranteed.

*Market to a third party* - in this case, resort to a "user interface" for the company's product catalog, user belonging to a third party (usually an Internet service provider or a bank). This single interface to several manufacturers of goods became known buyers, be attached to frequently accessed information channels (e.g. one button to access the most popular electronic journal).

*Virtual Communities* - The most important virtual community is given by its members (customers or partners) who add their own information over a basic environment provided by the company. Each member may offer for sale or may send requests for purchase of goods or services. Virtual community membership requires payment of a fee.

Value services provider for e-commerce channels - service providers specialized in specific functions such as providing logistics, electronic payment or expertise in production and inventory management. Payment of these services is based on rates or a percentage rate.

*Collaboration platforms* - platforms for collaboration include a set of tools and an information area for collaboration between companies.

Brokerage information's and other services - There were a lot of services that provide customers classified catalogs on your profile, business sales, investment advice, Consulting. A special category is a specialized trust services provided by certification authorities or electronic notaries. The ways for conducting electronic commerce transactions are usually accepted by all participants in the act of commerce. These rules called protocols for electronic commerce [3].The commercial transaction has three phases: negotiation, payment and delivery. Most of the literature focuses on the payment phase and, particularly, on how to ensure the security of electronic payments. The most important electronic payment protocols are: SET (Secure Electronic Transaction); SNPP (Simple Network Payment Protocol), IBS (Internet Billing Server)[5].

**Web Page - Business Card [13] obligatory**

In general, multinational companies allocate huge budgets to promote both offline and online. A website exposes the company to an audience of local, national and international, 24 hours a day, 7 days a week. A website is a simple and cheap method to communicate with customers and business partners. Unlike printed materials, a website does not run out, instead, it can be updated at any time with a new content. Customers can be directed to the website, where you shall purchase products or services are charged[7].

The publication on the World Wide Web is fast becoming a means of communication for all companies. Setting up a Web site is one of the best and most affordable ways to ensure a firm presence in the marketing and let you know the world that exists.

The Internet can help a company:

- *to become known*. A website can help a company to

be known and sell more products and services. A Web presence not only provides a new opening to customers, but they offer a way to get acquainted with the company without being bullied. Many people hate to call for information on a company and ask to see a sales representative. With a website, potential customers can obtain information on the conditions set by them.

- *reduce costs*. The Internet helps to lower costs of serving customers, whether new or current staff reduction.

- *to improve the customer service.* A website can be used to improve assistance to customers. Web sites are often used to help customers find the desired product, to remedy problems, to contact the company or even find helpful tips on how to make something better. And information is available 24 hours from 24, 7 days in 7. Many companies place on their website information about they product and services, useful advice and guidance to remedy any defects, and answers to frequently asked questions, so that customers and potential customers have access to information's without the need to call the company .

- *to enhance the company image.* Better informed and more knowledgeable in specific problems, the consumers started to consider normal for companies to have a Web site. For many companies, a Web site is a symbol of legitimacy and often, consumers, suppliers and others will want to do some research about the company, before entering into a business.

The internet development at the global and national level led to huge numbers of websites and web pages made in a business professional or less professional, according to the creator.

If at the beginning of the Internet phenomenon, the number of sites was reduced, the extent to which this phenomenon took it, year after year, made the number of sites to increase exponentially. Increasing number of sites determined by Web developers to create a true industry to meet customer requirements are becoming more demanding and more numerous.

The web page is a specific format of the Internet, which allows you to display information as text, sound and image, but also is an interface through which visitors can navigate within a site, generally on the Internet. A Web page is made easy, but creating a site requires more experience, training and use of specialized programs in graphic processing. In designing a professional site works in general, a team of four to five persons in charge of graphic design, database creation and promotion site.

The complex sites requiring and advanced programming, namely the creation of applications query and database management. Time allotted for completion of a site varies depending on complexity, but also the number of people involved in developing it. For a complex site requires about four weeks, during which determine the details, dimensions of the project and are effected.

To be on the Internet, a web hosting needs on a server connected to the Internet. In generally the companies have their own server, but Internet providers are made available, fee, space needed. Prices for locating the site on a server Internet service provider varies from 3 to 50 dollars a month, sometimes more, depending on the complexity of the site, the space on a server, the rates charged by each provider and, sometimes visits the site. Also, the site needs an address to be accessed and viewed. This address is composed of a coding standard (http://www.ro) and domain name (name, company, etc..). Sites can vary greatly. Depending on the purpose can be achieved: business presentations, online services, electronic stores, search engines. Mode can be accessed free at most sites, or pay. There are addresses for certain categories of customers: B2B (market "Business-to-Business" means the market transactions between firms), end-user and age categories.

A business like production sites are still growing quite difficult and is performed on uncertain ground, due to reduced orders and lower prices imposed by the Romanian market, but in time, the web site will become a book of access required. For the best business value requires a team with extensive experience and presence on the Web, which can provide more opportunities to promote effective address. Relations with other team owners of sites with advertisers on the Web, is an important advantage to exchange banners or links between sites.

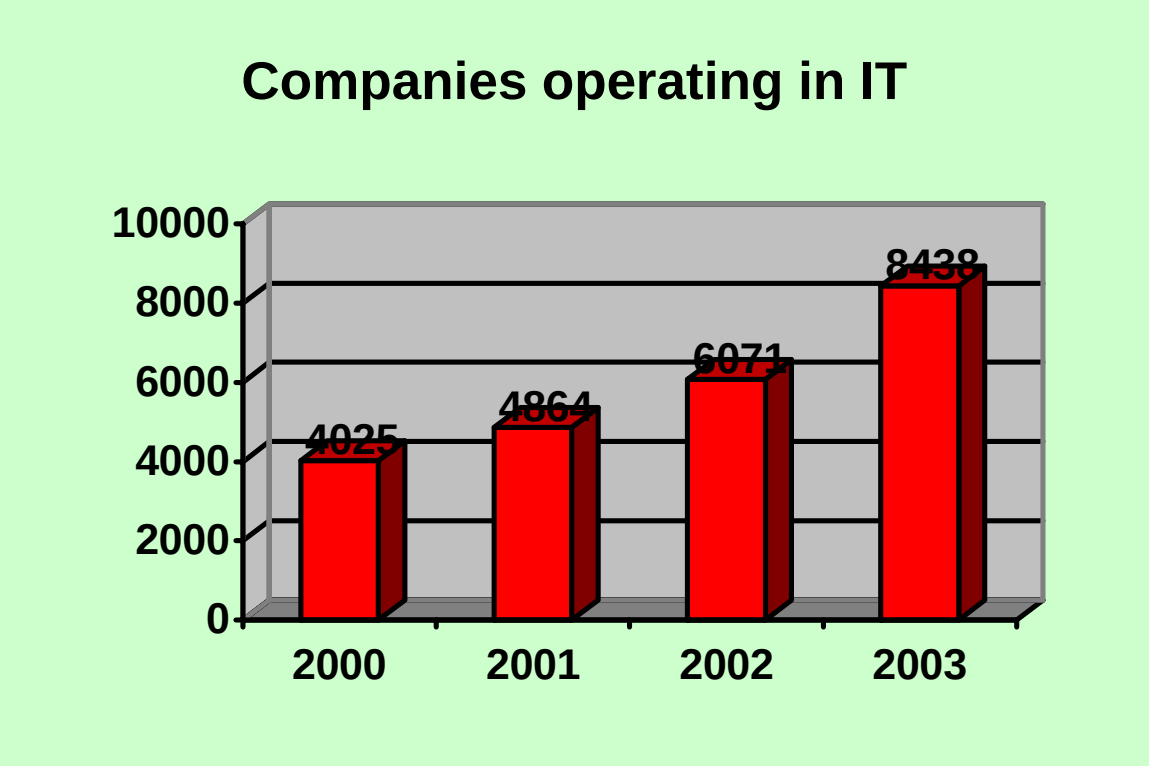

Fig. 1 Domenii.ro ROMANIA
Source-Site Ministry of Communications,
www.mcti.ro

Terminations for countries (. Fr,. Uk, hk,. En-etc) are unique, and acquired the name of a person or entity.

In Romania, it is used mainly areas. "ro", exclusively provided by the regulator of fields, Romanian National Computer Network (RNCN) at a purchase price of $ 61 including VAT. The ". Com" to buy the "life", unlike fields. Com /. "Net" for which payment is made annually at about 30-35 dollars and owned by the beneficiary.

If the RNC were recorded in March 1993 only two names ending in ". ro", their number began to grow rapidly in 1995, due to adopt Telecommunications Act, which authorized value-added services, reaching the end of 2004 to 68.000de addresses ending in ".ro. Telecommunications Act led to ISPs (Internet Service Provider ISP), forming thus a market in Romania for Internet. The number of companies in the Internet in 4026 doubled from 2000 to 8438 in 2003.

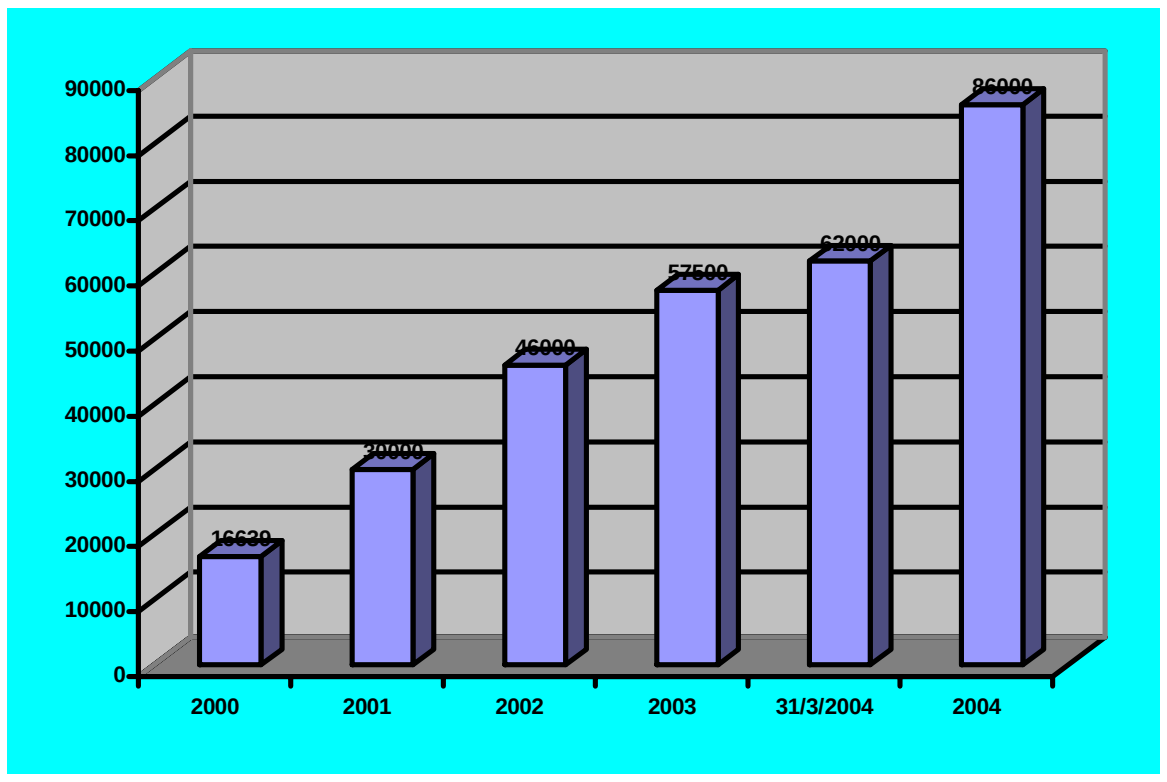

Fig. 2 Companies operating in IT

Source-Site Ministry of Communications, www.mcti.ro

*Particularly case of accounting for intangible assets internally generated. The costs of development and operation of websites*

**International Approach**

Given the substantial costs incurred by many companies for the creation and operation of websites, it became necessary to develop an accounting standard on accounting treatment of such transactions. In this regard was published in March 2002 SIC 32 - Intangible Assets-Costs associated with creating websites.

According to SIC 32 for costs associated with developing and maintaining websites are internally generated intangible assets falling under the governance of International Accounting Standard 38 - Intangible assets [12].

According to it, to be capitalized, the costs must be related expenditure "development phase" as described above. Must be strictly complied the conditions with the qualification phase of development and the ability to generate future economic benefits[10].

Thus, if a company will develop a website primarily for advertising and promoting its products and will not likely that this site will generate future economic benefits from its use, then these costs will be recognized as expense in the results( the profit and loss account) in period which they were made. On the other hand if your site will be used to take orders for delivery, the probability to generate future economic benefits are obvious and the costs will be fixed in accordance with IAS 38.

SIC 32 - Intangible Assets-Costs associated with the creation of websites provides a description of the phases of developing a website. They are:

**Planning**

This phase includes the initiation of feasibility studies, defining objectives and specifies, evaluating of the alternatives and selecting preferences;

| PLANNING | ACCOUNTING TREATMENT |
|---|---|
| - Initiation of feasibility studies (cost of salaries of the specialists who prepared the feasibility studies,etc.)<br>- Defining hardware specifications (configuration PC and other hardware components) and software (software used, etc.)<br>- Evaluation of alternative products and suppliers;<br>- Selecting preferences | The costs of the planning phase are not recognized as intangible assets. They are specific to the "research phase" and are recognized according to IAS 38, as spending in that period took place. |

**Application development and infrastructure**

This stage includes operations related to the domain name, purchase hardware components and operating software, installing developed applications and testing reliability.

| APPLICATION DEVELOPMENT AND INFRASTRUCTURE. | ACCOUNTING TREATMENT |
|---|---|
| - Acquisition hardware components | For this operation will apply the requirements of IAS 16 – Tangible assets. Hardware components will be recognized as tangible assets and will be assessed in accordance with IAS 16. |
| - Develop operating software (the software regarding operating system, software for creating and managing the site)<br>- Develop application code (the salaries of specialists)<br>- Installing the software developed by the web server (cost of salaries of specialized staff, consumable materials:<br>- Reliability testing | If the expense can not be directly attributed to preparing the website for operating it will be recognized as a period cost. If the expense can be directly attributed to preparing the website for operating in the manner intended by management and the website meets the criteria for recognition under IAS 38, then the costs will be capitalized (recognized as an |
| | intangible assets. |

**Develop graphic design**

At this stage is included the work of design of how the presentation of web pages.

| GRAPHIC DESIGN OF WEB PAGES. | ACCOUNTING TREATMENT |
|---|---|
| -the design graphic presentation of how the presentation of web pages<br>- the choice of format page<br>- the choice of colors, etc | At this stage the afferent costs should be recognized as intangible assets. However, if the expenditure can not be directly attributed to preparing the website for operating in the manner intended by management and are not satisfied the requirements of IAS 38, they will be recognized in the results account of the expends of period. |

**Develop the content of website**

This includes creating, purchasing, training and rescue of the information.

| DEVELOP THE CONTENT OF WEBSITE | ACCOUNTING TREATM |
|---|---|
| - Creating, acquitting, preparing information (identification tags, links, etc.).<br>- Saving information either as text or as graphics on the website before the website development;<br>- Content Information:<br>- products or services offered by the entity<br>- information about the company<br>- Sections that subscribers access<br>- contact details and correspondence | 1. If the website content is developed with the purpose of advertising and promotion of products and services, then costs associated with these activities will be classified as period costs.<br><br>2. If expenditure generated in this phase are directly attributable to preparing the website for operating in the manner intended by management and the website meets the conditions for recognition under IAS 38 (ability to generate future economic benefits and can be rated as reliable) will then be recognized as intangible assets. |

After completion the stage of development of the website begins the operational phase. In this phase, the entity maintains and develops software, infrastructure, graphics and content of the website. Expenses incurred during the operating cost should be recognized as a period in which they occurred, unless it met the recognition criteria set out in Standard. Specify the activities of this phase are listed below [13]:

| THE OPERATIONAL PHASE THE WEBSITE | ACCOUNTING TREATMENT |
|---|---|
| - Update graphics and content review<br>- Registration the website through search engines<br>- Adding new functions, features and content (links to other sites, etc.)<br>- Securing information<br>- Analyze the use of the site (count the number of visitors)<br>- Review how to access safety | The costs of these activities, if fulfill the conditions for recognition under IAS 38, will increase the value of the website. If you do not meet the conditions to be fixed will be recognized as expenses in the period in which they occurred. |

Other costs such as general administration costs, with training of employees on the use of the website will be recognized as expenses in the period indifferently if made during the development or phase of operation.

**Recognizing further websites**

Will apply the rules according to International Accounting Standard 38 - Intangible assets. Due to the technical characteristics and function of a website, some issues need to be nuanced.

Subsequent expenditure relating to the improvement or maintenance of their website has more to maintain the future economic benefits arising from the website, than to increase these benefits. This aspect will be further costs rarely restrained [7].

Another aspect to be shaded is about life. Because culminant progress in the field, websites likely to be very quickly outdated in terms of technology. Thus, entities must estimate the useful life on a prudent basis.

To further evaluate recognition initial moment, it is difficult for an entity to apply the revaluation model. To apply the revaluation, fair value should be estimated by reporting to an active market, or website is unlikely to have an active market. To these considerations is the best implementation model based on cost.

## B. National approach

National regulations do not provide rules or recommendations regarding the accounting treatment of the creation and development of websites. In the absence of legal regulations, there may be two options on the accounting treatment of the creation and development of websites.

The first variant involves the classification of costs by their nature, and in research expenditure and development expenditure. Costs of research will be rec-ognized as expenses in the period, while development expenditure will be restrained.

The second variant implies recognition of the cost of production, applying the same accounting treatment of software created in the enterprise.

The accounting treatment of intangible assets is governed by Order of Ministry of Public Finances. 1752 / 2005 for approval of accounting regulations with European Directives, as amended and supplemented by OMEF nr.2374 / 2007. Throughout the paper any reference to the Order of Ministry of Public Finances. 1752 / 2005 will be made by words with "national rules" [9].

National accounting rules defining identifiable intangible assets as assets, non-monetary, non-material support and held for use in the production or supply of goods or services to be leased to third parties or for administrative purposes.

As can be seen in the definition given by the national rules for intangible assets is introduced the identifiable nature of the asset, but without further details on identifying characteristics.

Compared to the international definition of referential IAS 38, the national standard includes in the definition of intangible assets and the purpose for which they are held by the entity.

**Conclusions**

International Accounting Standard 38 - Intangible Assets, requires for reporting entities presenting detailed information in the notes of the annual financial statements. Importantly, the standard does not require the presentation of comparative information on reconciling accounting amounts to the beginning and end of financial exercise. This exemption from presenting comparative information is not a departure from IAS 1 - Presentation of financial statements requires the principle of comparability of data reported. The paragraph 36 of IAS 1 - Presentation of financial statements is a clear indication that if standard enables information not be presented as comparative, then this requirement can not be met.

For each class of intangible assets, after the entity has made a distinction between intangible assets internally generated and other intangible assets, must be submitted by the following information in annual financial statements:

- the terms of life, if they are determined or undetermined, the depreciation rates used;
- depreciation methods used;
- gross carrying amount, accumulated depreciation and accumulated impairment losses, both at the beginning and the end;
- changes the carrying amount of intangible assets between the beginning and the end results of: inputs intangible assets (purchase or production of its own), revaluation, impairment or loss of depreciation losses, depreciation recognized during the reporting period, other changes;
- for intangible assets measured using the revaluation model, it is usually effective date of revaluation, the carrying amount of assets which would be their book value using the cost evaluation model, the revaluation surplus, the methods and significant assumptions applied in estimating values fair value of intangible assets;
- for intangible assets whose life was estimated as undetermined, will be presented: the book value and the reasons which make it has unlimited life and significant factors used for this estimate.

The contractual commitments for acquisition of intangible assets. The entity will present any restriction on the rights they have on restraint intangible entity and details of assets subject to security;

- for intangible assets acquired through a government grant and initially recognized at fair value, the entity will initially recognized fair value, book value and subsequent recognition method (method of cost or reassessment method);
- the aggregate amount of costs as research and development recognized at costs during the period;
- the descriptions of the property fully depreciated but still used. The entity is encouraged but not required to present this information;
- the information's on all intangible assets used by the company, but not recognized as assets in the balance sheet as it meet the criteria for recognition.

The economic entities often spend their resources or to attract acquisition debt, development, maintenance or expansion resources intangible assets such as scientific or technical knowledge, design and implement processes and systems, including among them and is web site. Because these elements can be recognized and valued in a accounting entity, must meet the definition of tangible assets and identifiable character, control over a resource and existence of future economic benefits. In case the item is gained through a combination of enterprises, it is a part of goodwill, recognized at data acquisition.

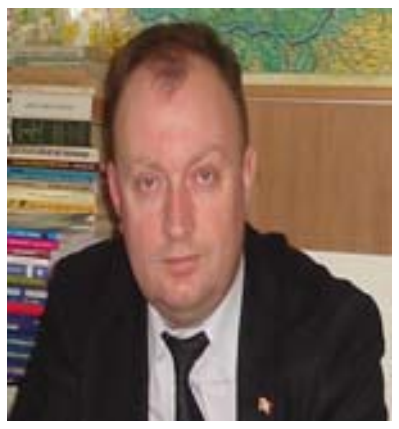

**Ionel Bostan** - **Prof. PhD-** Stefan cel Mare University, authors 5 speciality books, over 10 papers published in journals rated by ISI Thompson and 30 scientifically papers published in the country and abroad at the International Symposiums or Conferences. He is a member in 10 international professional organizations and scientifically referee in the editing committee of 2 journals rated by REPEC, Socionet, Research Gate etc, and within the scientifically committee of the WASET.

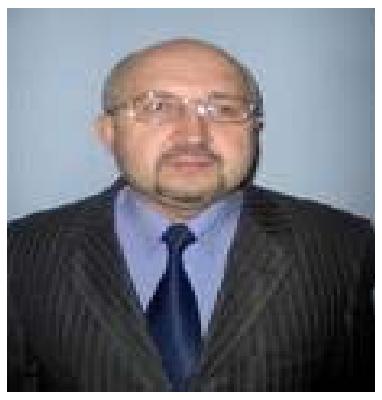

**Dorel Mates**, **Prof. PhD** - West University of Timisoara is a member in 6 academies and professional organisations, author and co-author of 7 books, over 50 scientifically papers, published abroad in specialty reviews or sustained and published within the international conferences and symposiums. He is a referee in the Science Committee of 2 foreign speciality journals and within the Science Committee of a foreign magazine.

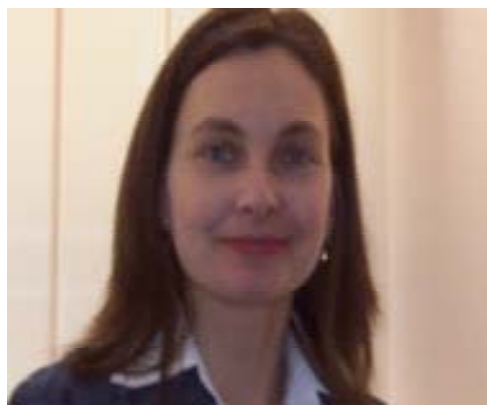

**Veronica Grosu**, **Asistent PhD** - Stefan cel Mare University of Suceava is a co-author of 3 specialty books, over 10 papers published in journals rated by ISI Thompson or within the international symposiums and conferences. She is a member in 3 international professional organizations.

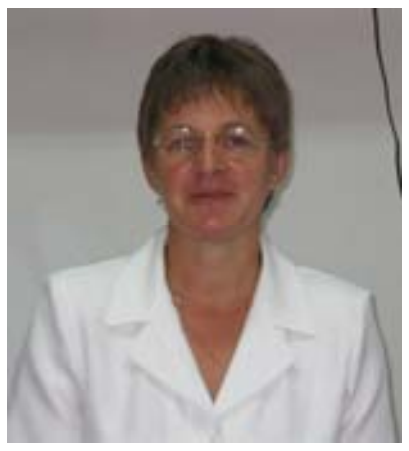

**Eugenia IANCU** Chair Lector of Informatics Desk, Economic Study and Public Administration Faculty, "Stefan cel Mare" University from Suceava. She is a doctorand at Technical Univesity from Timisoara. She's experienced in research contracts, she's part of the research team in 9 contracted projects, of which 4 are finalized and 5 are current.